\documentstyle[12pt]{article}
\advance\textheight by \topskip
\begin{document}

\rightline{SU-ITP-93-34}
\rightline{MSUHEP-93-22}
\rightline{UCSD/PTH-93-44}
\rightline{Dec 1993}

\vspace{.8cm}
\begin{center}
{\large\bf Lepton-antilepton Annihilation in the Massless Schwinger Model\\}

\vskip .9 cm

{\bf Jin Dai
\footnote{E-mail address: dai@higgs.ucsd.edu} }
 \vskip 0.1cm
Department of Physics 0319,
University of California, San Diego, \\
9500 Gilman Dr., La Jolla, CA, 92093-0319

{\bf James  Hughes
\footnote{E-mail address: hughes@msupa.pa.msu.edu} }
 \vskip 0.1cm
Physics and Astronomy Department,
Michigan State University, \\
East Lansing, MI  48823

{\bf
Jun Liu
\footnote{E-mail address: junliu@scs.slac.stanford.edu} }
 \vskip 0.1cm
Physics Department,
Stanford University,
Stanford, CA 94305

\end{center}

\vskip .6 cm
\centerline{\bf ABSTRACT}
\vspace{-0.7cm}
\begin{quotation}

We evaluate the next-to-leading order radiative corrections to
a process in the massless Schwinger model that is the analogue
of the $e^+e^-$ inclusive annihilation into hadrons process in
QCD.  We
then carry out an asymptotic expansion of the calculable exact
result and find that it agrees with the perturbative answer.
However, we also find that higher orders in the asymptotic
expansion involve terms that depend on the logarithm of the
strong coupling constant.

\end{quotation}

\normalsize
\newpage

\section{Introduction }

The most straightfoward process to analyze in QCD is the inclusive
electron-positron annihilation into hadrons. The leading order
QCD calculation confirms the charge and color assignments of the
quarks, and gives a rough measure of the heavy quark masses.
Further, a comparison of the next-to-leading order (NLO) calculation
with experiment can in principle provide a direct measurement of the
running of the strong coupling constant.

In the perturbative QCD calculations, the final states that are summed
over consist of colored quarks and gluons.  It is assumed that the
subsequent hadronization of these states affects the cross section
only by a factor whose difference from unity is suppressed by
powers of $m_h^2/q^2$, where $m_h^2$ is the squared hadronic mass
scale and $q^2$ is the invariant squared mass of the incident
$e^+e^-$ pair.  So far, this assumption has not been testable
because of the uncertainties in the experiments.

In this paper we analyze the analogous process in 1+1 dimensional
massless quantum electrodynamics, ${\rm QED}_2$.
The model exhibits both
confinement and asymptotic freedom and yet it is exactly solvable
\cite{sch,lowenstein,coleman,suss}.
Thus we may compare analytically the perturbative calculations with
the exact results and study the effects of hadronization.
What we find is an agreement between the NLO perturbative
cross section and the corresponding term in the asymptotic
expansion of the exact result
(both of order $g^2/q^2$ where $g$
is the ${\rm QED}_2$
coupling constant).  However, at higher orders in
$g^2/q^2$, we find  in the asymptotic
expansion  additional terms
that depend also on $\ln g^2/q^2$.  We did not expect that
perturbation theory could generate these terms and
we instead
associate them with the hadronic mass logarithms.

\section{Exact and Perturbative Cross Sections}
We calculate the analogue of $e^+e^- \rightarrow {\rm hadrons}$
in the following extension of the massless Schwinger model \cite{suss}.
The Schwinger model fermion, $\psi$, is
 Yukawa-coupled to a massless scalar field,
$\phi $, (our ``photon'').\footnote{ All of our amplitudes
will involve a single factor of the momentum space propagator of
this scalar field so we do not need to worry about
the nonexistence of massless scalar fields in $1+1$
dimensions.  Giving this scalar small
mass only effects the results by an irrelevant factor.}
A massless fermion $f$ (our ``lepton'') with zero ${\rm QED}_2$
charge is  also Yukawa-coupled to $\phi$
and for simplicity we will take both
fermions to have Yukawa charge $e$.  The full Lagrangian is
\begin{eqnarray}
L={F^2}/4+\bar{\psi}(i\not{\partial}+g\not{A})\psi\label{lagran}
+\bar{f}i\not{\partial}f+1/2\phi\Box\phi
+e(\bar{\psi}\psi+\bar{f}f)\phi.
\end{eqnarray}
So $g$ is the analogue of the ${\rm QCD}_4$ strong coupling constant and
$e$ is the analogue of the electromagnetic coupling.
We will confine all of
our calculations to the lowest order in $e$ and compare
the exact (all orders in $g$) result with the result of the
perturbation expansion
in $g$.
The exact results may be calculated using the
dual realization of the Schwinger model in terms of a free
scalar, $h$, of mass $m_h=g/\sqrt {\pi}$ and the bosonization formula
\footnote{The prefactor $c$ is a normal ordering dependent constant
and it equals $c=\frac{g \gamma }{(2\pi )^3/2}$ when the
normal ordering mass equals $m_h$.}
\begin{eqnarray}
\bar{\psi}\psi=c:\cos(2\sqrt{\pi}h):. \label{boson}
\end{eqnarray}

The process that we consider is then $f\bar f$ annihilation
$$
f(k_1)+{\bar f}(k_2)\rightarrow X,
$$
where where the momentum assignments of the fermions are given in the
parantheses
and the final states $X$
consist of quanta of the scalar field $h$ because the massless Schwinger
Model is equivalent to the free scalar free theory.
The cross section at leading order in $e^2$, but all orders in
$g^2$, is \cite{suss,jun}
\begin{eqnarray}
\sigma
=\frac{e^4}{q^4}\frac{1}{2}R(q^2).
\label{exactcross}
\end{eqnarray}
where $q=k_1+k_2$ is the energy of the virtual scalar photon,
and $R(q^2)$ is defined by
\begin{eqnarray}
R(q^2)=(R_{+}(q^2)+R_{-}(q^2))/2
\end{eqnarray}
and
\begin{eqnarray}
R_{\pm}(q^2)=c^2\int d^2x \exp(iqx) \exp (\pm 4\pi\Delta_{m_h}(x)),\label{rpm}
\end{eqnarray}
Here $\Delta_{m_h}(x)$ is the Wightman function for the free massive
scalar field $h$ and is calculated to be
\begin{eqnarray}
\Delta_{m_h}(x) = \int \frac{dp}{\sqrt{p^2+m_h^2}} \exp (ip\cdot x)
= \frac {1}{2\pi}K_0(m_h\sqrt{-x^2+i\epsilon x^0}).\label{deltaK}
\end{eqnarray}
and $K_0$ is the spherical Bessel function.

The parton model calculation of $f{\bar f}$ annihilation
is performed perturbatively in
$g$  where the final states that are summed over consist of
quanta of $\psi$ and of $A^{\mu}$.
The Infra-Red (IR) collinear divergences that afflict
${\rm QCD}_4$ also occur in
${\rm QED}_2$, but in $1+1$ dimensions as power
singularities rather than the logarithmic singularities
of
${\rm QCD}_4$.  We choose to regulate these
singularities by
temporarily giving the $A^{\mu}$ field a mass, $m_g$.
Then, as in
${\rm QCD}_4$, the KLN theorem
\cite {KLN} ensures that the inclusive
cross section, summed over possible
$\psi$ and of $A^{\mu}$ final states, will be finite as
$m_g \rightarrow 0$ at each order in perturbation theory.
At leading order in $g$ the only process that contributes
is
$$
f(k_1)+{\bar f}(k_2)\rightarrow
\psi(p_1)+{\bar\psi}(p_2),
$$
and the cross section is easily calculated to be
\begin{eqnarray}
\hat{\sigma}^0=\frac{e^4}{2q^4}.
\label{annx}
\end{eqnarray}

As in ${\rm QCD}_4$, the NLO $g^2$ corrections come from
the interference of the leading order
process with its one loop corrections,
virtual gluon corrections, $d\hat{\sigma}_v^1$,
and gluon bremstrahlung from one of the outgoing
quarks, $d\hat{\sigma}_b^1$.

First consider
the virtual corrections.  They consists of virtual correction to the
vertex and self energy corrections of quarks.
One of the reasons that we chose to regulate the IR collinear
divergence by giving the gluon a mass is that in $1+1$
dimensions the one loop fermion self energy graph vanishes.  This
is because the amplitude is proportional to
\begin{eqnarray}
{\rm Tr}[\gamma_{\mu}(\not{p}-\not{k})\gamma_{\mu}]
=(2-d){\rm Tr}[\not{p}-\not{k}]=0,
\label{self}
\end{eqnarray}
where $p^{\mu}$ is the fermion momentum and $k^{\mu}$ is
the loop gluon momentum.  Note that this trace
vanishes For $d=1+1$ spacetime.  So we only need to consider the
vertex loop correction of a gluon exchange between the outgoing
quarks.  A direct calculation gives the invariant amplitude
for this process to be
\begin{eqnarray}
    M_v = e^2 g^2 \int d^2k \frac{ \bar{u}(p_1) \gamma^\mu
        (\not{p_1}+\not{k})(\not{p_2}-\not{k}) \gamma_\mu v(p_2)}
        {(p_1+k)^2 (p_2-k)^2 (k^2-m_g^2)}.
\end{eqnarray}
Using the formula $\gamma _{\mu}\not{a}\not{b}\gamma ^{\mu}
=2\not{b}\not{a}$ for $1+1$ dimensions and standard Feymann integration
aproach one can prove that
\begin{eqnarray}
  & & \int d^2k \frac{ \bar{u}(p_1) \gamma^\mu (\not{p_1}
        +\not{k})(\not{p_2}-\not{k}) \gamma_\mu v(p_2)}
        {(p_1+k)^2 (p_2-k)^2 (k^2-m_g^2)}   \nonumber \\
  &=& -2\pi i\sqrt{2E_1 2E_2} (m^2_g+q^2)
      \int^1_0 dx \int^{1-x}_0 dy \frac{1-x-y}{(1-x-y)m^2_g -xy q^2}
              \nonumber  \\
  &=&  -2\pi i\sqrt{2E_1 2E_2} \frac{1}{q^2 \beta} \ln\beta ,
\end{eqnarray}
where $\beta \equiv m_g^2/q^2$.

The interference of this one-loop amplitude with the leading
amplitude then gives the following order $g^2/q^2$ cross section:
\begin{eqnarray}
   \hat\sigma_v^1 = \frac{e^4}{2\pi q^4} \frac{g^2}{q^2}
                    \frac{1}{\beta} \ln{\beta} .
\end{eqnarray}
Next consider gluon bremstrahlung
$$
f(k_1)+{\bar f}(k_2)\rightarrow
\psi(p_1)+{\bar\psi}(p_2)+A^{\mu}(p_3),
$$
The invariant amplitude for this process is
\begin{eqnarray}
    M_b = -ie^2g^2 \frac{1}{q^2} \bar{u}(p_1)
         \left[ \not{\epsilon} \frac{\not{p_1}+\not{p_3}}{(p_1+p_3)^2}
         -\frac{\not{p_2}+\not{p_3}}{(p_2+p_3)^2}\not{\epsilon}
         \right] v(p_2) .
\end{eqnarray}
Where $p_3$ is the ``gluon'' momentum and $\epsilon$ its polarization.
It is a simple exercise in kinematics to derive the corresponding
cross section
\begin{eqnarray}
  d\hat\sigma^1_b = \frac{e^4}{2\pi q^4} \frac{g^2}{q^2}
                    \frac{1}{\pi\beta} \frac{1}{1-x_1}dx_1
\end{eqnarray}
where $x_1\equiv 2p_1 \cdot q/{q^2}$.

The corresponding gluon-emmission differential cross section in
${\rm QCD}_4$
is finite
(see, for example, \cite{field}), and the
collinear divergences arise only upon
the integration over phase space.  Indeed,
apart from the phase space points where the gluon is collinear with
one of the quarks, the differential cross section describes
the probability of a 3-jet event which is observable and so
the corresponding cross section must be finite.
In contrast, in $d=1+1$ the emitted gluon is
forced to be collinear, there are no jets, and so the differential
cross section is infinite for $\beta = 0$. This is fine
because the only physical observable
is the total cross section which, as we will see, is finite.

The maximum value of $x_1$ is $1-\beta $,  So integrating the
differential cross section gives the NLO gluon emission cross
section to be
\begin{eqnarray}
  \hat\sigma^1_b = -\frac{1}{2\pi}\frac{e^4}{q^4}\frac{g^2}{q^2}
                    \frac{1}{\beta} \ln\beta
\end{eqnarray}
The result for the total cross section including
including all of the order $g^2/q^2$
perturbative corrections is thus
\begin{eqnarray}
  \bar\sigma = \hat\sigma^0 +\hat\sigma^1_v + \hat\sigma^1_b
= \hat\sigma^0       \label{partcross}
\end{eqnarray}
That is, the loop correction, $\sigma ^1_v$ exactly cancels
the bremstrahlung correction, $\sigma ^1_b$.  The cancellation
between these cross sections of the terms that are singular
in $m_g$ is not a surprise -- it is understood to be a consequence
of the KLN theorem \cite{KLN}.  However, it is surprising that,
unlike ${\rm QCD}_4$  there is no
finite residue.  To compare this NLO result with the exact
result in equation (\ref{exactcross}), we will next expand the
functions $R_{\pm}(q^2)$ in
powers of $g^2/q^2$.

\section{Asymptotic Expansion of the Exact Result}
The functions $R_{\pm}(q^2)$ which describe all of the interesting
dynamics are compactly given in terms of 2-dimensional
Fourier transforms (FT) of the exponential of the Wightman function
in equation (\ref{rpm}).
It turns out that the evaluation of the asymptotic expansion of the
$R_{\pm}(q^2)$ may be reduced to a straightfoward exercise involving
the 1-dimensional FTs of generalized functions \cite{gelfand}.
We will describe briefly the steps involved in the expansion
bescause the calculation is rather involved.
We will then evaluate explicitly the first few terms of the expansion
and demonstrate the occurrence of the advertised
$\ln g^2/q^2$ term at order $g^6/q^6$.

We utilize the fact that when $q \rightarrow \infty $, the FT of
a function is dominated by its singularities on the real axis, in this
case, at the origin. So all we need to do is to  pick out the power
and logarithmic singularities of the function
$\exp\left[\pm 4\pi \Delta_{m_h}(x)\right]$, and organize them according
to their power.
All the analytic terms will drop out,
as their contributions are exponentially
small.

The first step in carrying out the asymptotic expansion of $R_{\pm}(q^2)$
is to expand the Wightman function that appears in the
argument of the exponential in equation (\ref{rpm}).
The expansion of  $\Delta_m(x)$ given in equation (\ref{deltaK})
is  (this is the standard expansion of spherical Bessel functions \cite{grad})
\begin{eqnarray}
  4\pi\Delta_{m_h}(x) = \sum_{k=0}^{\infty}\frac{1}{2^{2k}(k!)^2}
  [m_h^2(-x^2+i\epsilon x^0)]^k(2\psi (k+1)-
      \ln [\frac{m_h^2}{4}(-x^2+i\epsilon x^0)]).\label{deltaexp}
\end{eqnarray}
This expansion appears in the argument of the exponential in
equation (\ref{rpm}). When the exponential is expanded and
the terms in the double expansion are grouped according to their
power dependence on
$m_h^2(-x^2+i\epsilon x^0)$.
All  we have to evaluate are
2-dimensional FTs of terms of a power times a logrithmic function.

Next, by using the equalities
\begin{eqnarray}
-x^2+i\epsilon x^0 = -2(x^+-i\epsilon)(x^--i\epsilon),
\label{factor}
\end{eqnarray}
and
\begin{eqnarray}
d^2x=dx^+dx^-,
\end{eqnarray}
where
\begin{eqnarray}
x^{\pm}\equiv \frac{1}{\sqrt 2}(x^0\pm x^1) .
\end{eqnarray}
The expansion of
$R_{\pm}(q^2)$ is
reduced to taking products over
the 1-dimensional FTs
of power and logrithmic functions of $x^+$ and $x^-$.
\begin{eqnarray}
\int_{-\infty}^{\infty}dx
    (x^{\pm}-i\epsilon)^n \ln^m(x^{\pm}-i\epsilon) \exp {ix^{\pm}p},
     \label{FT}
\end{eqnarray}
where $n$ and $m$ are integers.

We have carefully kept track of the regularizations which are
need to make these FTs well-defined. The techniques
 are discussed in detail in the
book by Gel'fand and Shilov \cite{gelfand}.

The FTs in (\ref{FT}) may all be generated by taking
derivatives with respect to $\lambda$ of the following FT equality given
in entries 25 and 26 in the Table of
Fourier Transforms in \cite{gelfand}:
\begin{eqnarray}
\int_{-\infty}^{\infty}dx
(x\pm i\epsilon)^{-\lambda}
\exp {ixp}=
\frac{2\pi \exp {(\mp i\lambda \pi /2)}}
{\Gamma [\lambda ]}p^{\lambda -1}_{\mp}.
\label{FTlamb}
\end{eqnarray}
The subscript $\mp$ for $p$ on the right-hand-side (rhs) of
(\ref{FTlamb})  defines generalized functions:
\begin{eqnarray}
p^{\lambda -1}_{\mp}=\vert p\vert ^{\lambda -1}\theta (\mp p) .
\end{eqnarray}
and $\theta$ is the usual step function.  To generate the desired FTs we
expand $\lambda$ in Laurent series the
terms on the rhs of (\ref{FTlamb}), collect terms with the same power
dependence, and take derivatives with respect to $\lambda$.  The result
is a series of FT of the functions
of powers of $x\pm i\epsilon $ times powers of the logarithm of
$x\pm i\epsilon $.  For example,
\footnote{ Here $\psi(n+1)$ is the logarithmic derivative of the usual
gamma function evaluated at $n+1$, so
$\psi(x)\equiv \frac{\Gamma '(x)}{\Gamma (x)}$.  Refer to the section
on the psi function in \cite {grad}.}
\begin{eqnarray}
   \int_{-\infty}^{\infty}dx
         (x -  i\epsilon)^n \ln (x -  i\epsilon)\exp {ixp}
   &=&\frac{2\pi}{i^n}
     \left\{ \delta^n(p)\left[\psi(n+1) - \frac{\pi}{2} \right]
        - \frac{2n!(-)^n}{p_{+}^{n+1}} \right\},
\end{eqnarray}
and
\begin{eqnarray}
   \int_{-\infty}^{\infty}dx
         (x -  i\epsilon)^n {\ln }^2 (x -  i\epsilon)\exp {ixp}
   &=&\frac{2\pi}{i^n}
     \left\{\delta^n(p)\left[\psi '(n+1)+\psi ^2(n+1)  -  i\psi (n+1)
          -\frac{7}{12} \pi ^2\right] \right.     \nonumber \\
   & &\left. +\frac{2n!(-)^n}{p_{+}^{n+1}}
    \left[\ln (p_{+}) -\psi (n+1) + i\pi /2  \right] \right\} .
\label{FTex}
\end{eqnarray}
The $\delta$ function terms drop out when $|p|$ is large.

The final result of these three steps described above - expanding the
integrand in the 2-dimensional FT definition of $R_{\pm}(q^2)$ in
powers of
$m_h^2(-x^2+i\epsilon x^0)$, factorizing each term into products
of 1-dimensional FTs over $x^+$ and $x^-$, and evaluating the
resulting FTs using the master result in equation (\ref{FTlamb}) -
is an asymptotic expansion of $R_{\pm}(q^2)$ in powers
of $g^2/q^2$.  The expansion can be taken out to arbitrary
order by applying these steps.  We have evaluated the terms out
to order $g^8/q^8$ and find that for $q^2\ne 0$ and
$q^{\pm}>0$:
\begin{eqnarray}
R_+(q^2)= 1 +\frac{1}{2\pi ^2}\frac{g^4}{q^4}
+\frac{g^6}{q^6} \frac{1}{\pi ^3} (12+4\ln [\frac{\pi q^2}{g^2}])
+ o(\frac{g^8}{q^8}),
\label{rplus}
\end{eqnarray}
and
\begin{eqnarray}
R_-(q^2)=
\frac{g^8}{q^8} \frac{1}{\pi ^4} (6.96-3.79\ln [\frac{\pi q^2}{g^2}])
+ o(\frac{g^{10}}{q^{10}}).
\label{rmin}
\end{eqnarray}

There are several points to make regarding the above expansion.
First, the
fact that the terms of order $g^2/q^2$ vanish in both of the
expansions in equations (\ref {rplus}) and (\ref {rmin}) implies that the
NLO term in the expansion of the exact cross section
for $f{\bar f}$ annihilation,(\ref {exactcross}), also vanishes.
This is in agreement with the
perturbative result, equation (\ref{partcross}).
Second, for either $q^+ <0$ or $q^- <0$ the functions $R_{\pm}(q^2)$
vanish.  Also, the
$R_{\pm}(q^2)$ contain delta function contributions concentrated
on the lightcone, $q^+ =0$ or $q^- =0$.  Basically, all the terms
in the FT's that are not delta functions depend on $q_+^{\pm}$, see
equation (\ref {FTex}).  Futhermore this property
of $R_{\pm}$ - that it is nonzero only in the foward lightcone -
persists at all orders in the expansion and can be traced
back to the boundary conditions on the Wightman function
$\Delta _m(x^2)$ that leads to the regularization
$x^2\rightarrow x^2-i\epsilon x^0$.
The third point is that the calculational steps outlined
above imply that the asymptotic expansions of the $R_{\pm}$ functions
have the form (again in the foward lightcone):
\begin{eqnarray}
R_+(q^2)= 1 +\sum _{n=2}^{\infty}
(\frac{g^2}{q^2})^n
(a^1_n + a^2_n \ln [\frac{q^2}{g^2}] +\cdot \cdot \cdot +
a^{n-1}_n \ln ^{n-2} [\frac{q^2}{g^2}])
\label{rplusgen}
\end{eqnarray}
and
\begin{eqnarray}
R_-(q^2)=
\sum _{n=4}^{\infty}
(\frac{g^2}{q^2})^n
(b^1_n + b^2_n \ln [\frac{q^2}{g^2}] +\cdot \cdot \cdot +
b^{n-2}_n \ln ^{n-3} [\frac{q^2}{g^2}]),
\label{rmingen}
\end{eqnarray}
where the $a$'s and $b$'s are $q$- and $g$-independent
numbers.  These expansions are all that we will need to know about
the $R_{\pm}$ functions in this and succeeding work.
For example, a review of the cross sections given in \cite {hugliu1}
together with the above result that the $g^2/q^2$ terms vanish
imply that the NLO contributions
to the cross sections must vanish.  We will use this result in
subsequent work
\cite {dhl}
to compute the NLO distribution and fragmentation
functions for the model defined above in equation (\ref{lagran}).

The final point is that, as advertised in the abstract, the asymptotic
expansion of the exact result for the cross section contains a
$\frac{g^6}{q^6} \ln[\frac{g^2}{q^2} ]$ term which comes from
FT of $\ln^2(x^2)$.  This term is a new kind of non-perturbative effect,
it is much more stonger than the common exponetially small
non-pertubative effect( e.g., instantons in ${\rm QCD}_4$ ) which are dropped
in this expansion. Maybe it associated with hadronization.  Note that
for large enough $q^2$ the log term will dominate the other term (that
we presume is calculable in perturbation theory) at order $g^6/q^6$.

\section{Conclusion}
Simple dimension counting in ${\rm QCD}_4$ distinguishes the
perturbative corrections according to their twist.  Thus, both
the higher twist corrections and, by assumption, the
hadronization corrections are suppressed by powers of
$1/q^2$.  The dimension counting is different in
${\rm QED}_2$ where the coupling $g$ has the dimensions
of a mass.  Here all corrections are suppressed
by powers of $g^2/q^2 \propto m^2/q^2$.  Nevertheless, the
NLO perturbative result agrees with the asymptotic
expansion at order $g^2/q^2$.  Further, the presence of
the $(g^2/q^2)^3 \ln g^2/q^2$ term in the expansion of the
exact result is a surprise from the point of view
of perturbation theory and poses a challenge to
calculate.

In the preceding paper \cite{hugliu1} the only property of the exact
result that was used was that $R(q^2)\rightarrow 1$ as
$q^2 \rightarrow \pm \infty$.  In this paper, the complete
asymptotic expansion of $R(q^2)$ is described and the first
few terms are calculated explicitly out to order $(g^2/q^2)^3$.
In subsequent work \cite{dhl} we will use this expansion to
extend the analysis in \cite{hugliu1} to the next-to-leading order.  That
is, to analytically check for process independence of
the parton distribution and fragmentation functions in the
massless Schwinger model.

\section*{Acknowledgements}
We have benefited from discussions with S. Brodsky, J. Bjorken,
M. Peskin, H. D. Politzer, and L. Susskind.
J.H. was supported during part of this work
by the US DOE under Contract No. W-7405-ENG-48(LLNL) and the Nuclear
Theory Grant No. SF-ENG-48.
J.D. was supported by DOE under grant DE-FG03-90ER40546. J.L. was
supported in part by the National Science Foundation grant PHY
89-17438.
\vskip 1cm


\end{document}